\documentclass[a4paper]{article}

\usepackage[pages=all, color=black, position={current page.south}, placement=bottom, scale=1, opacity=1, vshift=5mm]{background}
\usepackage[margin=1in]{geometry} % full-width

\usepackage{amsmath}
\usepackage{amsthm}
\usepackage{amssymb}
\usepackage[version=3]{mhchem}

\usepackage[utf8]{inputenc}
\usepackage{hyperref}
\hypersetup{
	unicode,
	pdfauthor={Author One, Author Two, Author Three},
	pdftitle={A simple article template},
	pdfsubject={A simple article template},
	pdfkeywords={article, template, simple},
	pdfproducer={LaTeX},
	pdfcreator={pdflatex}
}

\usepackage[sort&compress,numbers,square]{natbib}
\theoremstyle{plain}

\theoremstyle{definition}

\usepackage{graphicx, color}
\graphicspath{{fig/}}

\usepackage{algorithm, algpseudocode} % use algorithm and algorithmicx for typesetting algorithms
\usepackage{mathrsfs} % for \mathscr command

\usepackage{lipsum}

\title{Motif based high-throughput structure prediction of superconducting monolayer titanium boride}
\author{Jason Yu$^1$\thanks{\texttt{morty.yu@yahoo.com}} \and Jihai Liao$^1$ \and Xiaobao Yang$^1$ \and Yujun Zhao$^1$ \and Yinchang Zhao$^2$}

\date{
	$^1$Department of Physics, South China University of Technology,Guangzhou 510640,People’s Republic of China\\%
	$^2$ Department of Physics,Yantai University,Yantai 264005,P. R. China \\[2ex]%
}

\begin{document}
\maketitle

\begin{abstract}
Two-dimensional boron structures, due to the diversity of properties, attract great attention because of their potential applications in nanoelectronic devices. 
A series of \ce{TiB_x} ($4\leq x \leq 11$) monolayers are efficiently constructed through our motif based method and theoretically investigated through high-throughput first-principles calculations.
The configurations are generated based on motifs of boron triangular/quadrilateral fragments and multicoordinate titanium-centered boron molecular wheels. 
Besides priviously reported \ce{TiB4} which was discovered by global search method, we predict that high symmetry monolayers \ce{TiB7} (Cmmm) and \ce{TiB9} (P31m) which are octa-coordinate and nona-coordinate titanium boride are thermodynamic stable. The \ce{TiB7} monolayer is a BCS superconductor with the transition temperature $T_c$ up to 8K.
The motif based approach is proved to be efficient in searching stable structures with a prior knowledge so that the potentially stable transition metal monolayers can be quickly constructed by using basic cluster motifs. 
As an efficient way of discovering materials, the method is easily extended to predict other type of materials which have common characteristic pattern in the structure.
\end{abstract}

% \tableofcontents
\section{Introduction}
Boron demonstrates the structural diversity in low-dimensional systems due to the multicenter bonds, from zero-dimensional clusters to two-dimensional layers. 
The most stable \ce{B12} cluster is a fragment of planar triangular grids\cite{Zhai2003}, which otherwise forms a highly symmetric icosahedron as bulk cluster\cite{Fujimori1999}. 
As the size of clusters increases, the hexagonal vacancies were found to be energetically preferrable\cite{Li2014,Piazza2014}, such as \ce{B30} and \ce{B36}. 
Meanwhile, the stable boron monolayers with hexagonal vacancies were theoretically proposed\cite{Tang2007,Yang2008}. Various kinds of distributions of vacancies in borophene cause a variety of novel properties, including topological property\cite{Feng2017}, superconductivity\cite{Penev2016,Zhao2016} and even forming the borophene semiconductor\cite{Xu2017}. 
In the recent experiments\cite{Feng2016,Liu2018}, boron monolayers were fabricated on the metallic substrates by the molecular beam epitaxy method, where the vacancy patterns could be neatly modulated by the type of substrate\cite{Wu2019} and the growth conditions\cite{Zhang2019}. 

The structural diversity and corresponding properties of boron nanostructures were able to be further enhanced by nesting the metal atoms\cite{Zhang2016,Zhang2014,Romanescu2011,Li2019,Simonson2010}. 
For instance, the band structure of quasi-monolayer \ce{TiB2} \cite{Zhang2014} was predicted to be characterized with anisotropic Dirac cones and the Fermi velocity of the material was as large as half of the graphene. 
As for the two-dimensional \ce{MoB4}\cite{Simonson2010}, it revealed a novel electronic structure which contains double Dirac cones near Fermi level with high Fermi velocity and superconductivity with transition temperatures ($T_c$) of 44.5K. 
Experimentally, various planar hyper-coordinate species of transition metal (TM)@\ce{B_n} with transition metal atom locating at the center of the boron wheel were synthesised \cite{Romanescu2011}, which might serve as building blocks for the construction of transition metal boride monolayer. 

Based on the first-principles calculations, the \ce{FeB6} monolayer was predicted to be stable with the semiconducting properties\cite{Li2016}, and the \ce{CrB4} monolayer was prediced to be ferromagnetic with the Curie temperature of 401K\cite{Li2019}. 
Transition metal elements exhibit rich magnetic properties because of their their unparied electrons in $d$-orbitals. 
Despite the fact that the superconductivity have been predict to be exist in two-dimensional \ce{Mo2B2}\cite{Yan2019} and Li-decroated boron monolayers\cite{Wu2016}, the superconductivity is still out of sight in the transition metal boron monolayer.

Generally, the polymorphism of boron monolayers attribute to the variety of properties, but at the same time it become a challenge to screen the candidates due to the complicated potential energy surface\cite{De2011}.
Combined with the global optimization with the first-principles calculaltions, extensive researches have been focused on giving the concentration and distribution of vacancies in boron monolayers\cite{Wu2012,Yu2012}, which might be synthesized on the certain metal surfaces. \cite{Zhang2015}
In our previous studies, we used the motif of triangular lattice with hexagonal vacancies to construct possible borophenes which base on the inductive conclusion that when the number of boron atoms less than four, the borophene became less stable\cite{Xu2018}. With this a prior knowledge, we explored all the structures through high-throughput first-priciple calculation and the semiconducting monolayers were successfully predicted\cite{Xu2017}. 
By using the similar strategy, the superconducting metal-boron monolayers are expected to be found based on the biased screening with proper motifs with the following two a prior knowledge a) transition metal and boron are were to form the stable planar clusters, b) the reported monolayer metal-boron materials were made up of hyper-coordinate metal-boron clusters and boron triangular grid fragments. 

In the paper, an efficient method is proposed to construct Ti@\ce{B_n} with the polygonal transition metal wheel motifs and triangular/quadrilateral boron motifs. Combining with the structue duplicate removal approach, we performed the high-throughput first-principles calculations on  \ce{TiB_x} ($4\leq x \leq 11$) monolayers, with the help of AiiDA \cite{Pizzi2016} to maintain the reproducibility of the computational process.  Besides \ce{TiB4} which is the same to the priviously reported one, we predict two new stable monolayers of \ce{TiB7} and \ce{TiB9}. Notably, the \ce{TiB7} monolayer is found to be superconducting with the $T_c$ of 8K. 
\section{Computational methods} \label{ss:method}

\begin{figure}
    \centering
    \includegraphics[width=8.7cm]{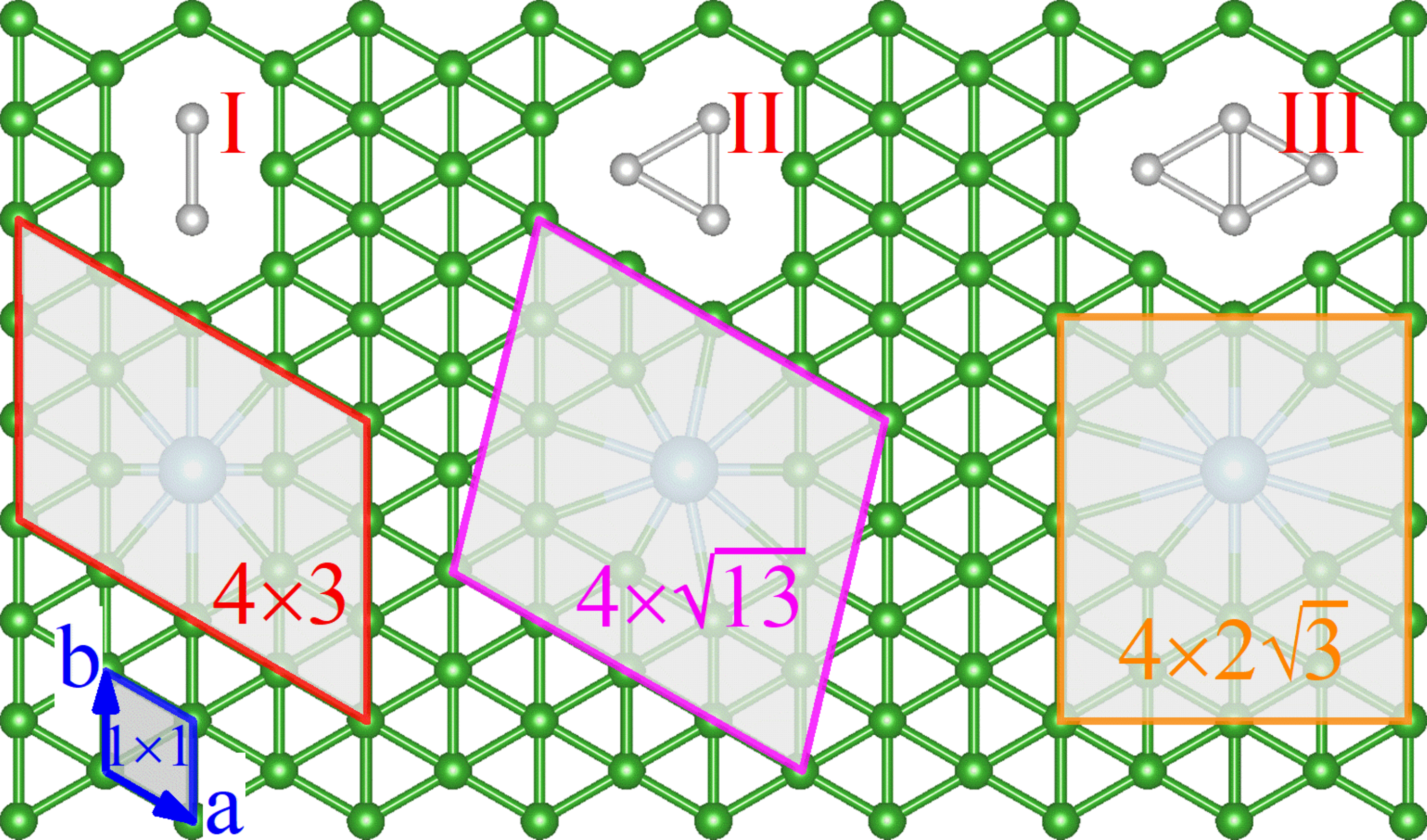}
    \caption{A schematic diagram of structure generation for monolayer titanium boride. A titanium atom substitutes two, three, and four boron atoms in type I, II, and III, respectively. The rhombus outlined by blue lines denotes the unitcell of 2D triangular lattice. Some supercells are also shown. The big sky-blue and small green spheres represent titanium and boron atoms, respectively. The gray spheres represent boron atoms to be substituted.}
    \label{fig:how-constructed}
\end{figure}

In the recently proposed \ce{TiB2} sandwich structure\cite{Zhang2014}, the vertical distance between titanium and boron network is 1.19 \AA. Similar sandwich structure have also been predicted to be stable in titaniuim boride and iron boride. \cite{Xu2017b}
Here our purpose is to search the stable monolayer transition boride. To form the monolayer transition metal boride, we need a boron network with transition metal embedded in. However, the transition metal such as titanium is too large to be embedded into the center of boron hexagonal ring to constructe the hexa-coordination motif. 
Therefore,  hyper-coordination larger than six will be necessary for the creation of Ti-B  monolayers. 
In order to sample the configurations efficiently in the phase space, we construct the possible candidates with specific motifs for the following reasons:
a) stable boron monolayers are the triangular lattice with vacancies and transition metal atoms serve as the electrons donor;
b) many hyper-coordinate transition boron clusters have been observed experimentally, which is considered to be the building blocks of the correspond monolayers.

The following steps were used to ergodiclly sample the configurations with the constrains mentioned above. 
First of all, the superlattice of boron lattice was created from extending the primitive of lattice which is shown in the Fig.\ref{fig:how-constructed} (the blue quadrilateral in the left bottom corner). The unique deravative superlattice with the unit cell sizes from 1 to 16 can be generated using Hermite Normal Form (HNF) matrices. \cite{Hart2008}
Each HNF matrix generates a superlattice of a size corresponding to its determinant $n$ which is also the number of the boron atoms of the lattice. 
There exists many HNF matrices with the same determinant, each creating a variant superlattice.
For example, the lattice in the Fig.\ref{fig:how-constructed} are three lattice with different HNF matrices.
Secondly, for each unique superlattice, we can substitute the two-body, three-body and four-body boron clusters by titanium atoms, where the type I, II and III correspond to the \ce{B2}, \ce{B3}and \ce{B4} clusters are shown in the Fig.\ref{fig:how-constructed}. 
Finally, based on the structural recognition, we removed the duplicate configurations in order to prevent repeated computation. 

{{}The first-principles calculations were performed using the Vienna ab initio simulation package (VASP).\cite{Kresse1996}
The projector-augmented plane wave (PAW) approach was used to represent the ion-electron interaction.\cite{Kresse1999,Blochl1994}
The electron exchange-correlation functional was treated using generalized gradient approximation(GGA) in the form proposed by Perdew, Burke and Ernzerhof(PBE).\cite{Perdew1996}
{}}
The energy cutoff of the plane wave was set to 350 eV.
The Brillouin zone was sampled with allowed spacing between k points in 0.2 \AA$^{-1}$, with Monkhorst-Pack k-points grid for high-throughput geometry optimization.
After high-throughput screening, the geometry optimization with high accuracy is performed for lowest ten structures in each composition.
In this process, the atomic position were fully relaxed until the maximum force on each atom was less than $10^{-2}$ eV/\AA.

% SC calculation method
{}{}To study the superconductivity of \ce{TiB_x}, the electron–phonon coupling(EPC) calculations were performed based on the density functional theory (DFT) and density functional perturbation theory (DFPT) implemented in QUANTUM-ESPRESSO package\cite{Giannozzi2009}, employing the projector augmented-wave (PAW) pseudopotentials\cite{Kresse1999,Blochl1994} with Perdew-Burke-Ernzerhof (PBE) exchange-correlation energy functional \cite{Perdew1996}. The plane-wave cutoff energies for wavefunctions and charge density were set to 60 Ry and 600 Ry, respectively. A Marzari-Vanderbilt cold smearing \cite{Marzari1999} of 0.02 Ry was used for the corresponding electronic self-consistent cycles. Phonon frequencies and EPC parameter were calculated with a phonon wave-vector mesh of $6\times6\times1$ and a dense k mesh of $24\times24\times1$, respectively.

% AiiDA part
Using Structures of Alloys Generation And Recogition (SAGAR) package\cite{sagar2019}, we have constructed  the possible configurations up to the  unit cell sizes up to 16 atoms, including 210, 98, 150 unique structures for type I, type II and type III substitution stratigies respectively. During the  high-throughput screening, the large number of simulations are involved and the complex sequence of logical steps are required in the study.
To ensure reproducibility, we use AiiDA \cite{Pizzi2016} as an open-source materials' informatics infrastructure to implement the calculation, keeping track of the full provenance of each calculation and results. All the calculation details is stored and exported to files which can be found in the materialcloud.

\section{Results and discussion}
Based on the  the high-throughput first-principles calculations, we systematically investigate the structural stabilities of \ce{TiB_x} ($4\leq x \leq 11$) monolayers in Section \ref{ss:structure}. The monolayers with superconductivity are discussed in Section \ref{ss:superconducting}, according to the electron–phonon coupling calculations.

\subsection{Structure stabilities of \ce{TiB_x} ($4\leq x \leq 11$) monolayers} \label{ss:structure}
Due to the remarkable difference of Ti-B and B-B bond lengths, titanium is considered to be unfavourably embedded in the 6-membered rings and forming hexa-coordinate wheels. In the recent study\cite{Qu2017}, the \ce{TiB4} monolayer was predicted to be a completely planar structure, in which B atoms form a network made up of 4-membered and 8-membered rings. Thus, the \ce{TiB4} monolayer was considered to be constructed with the motif of \ce{TiB8} wheels, where all titanium atoms were all embedded in the boron octagons. 

\begin{figure}[htb]
    \centering
    \includegraphics[width=8.3cm]{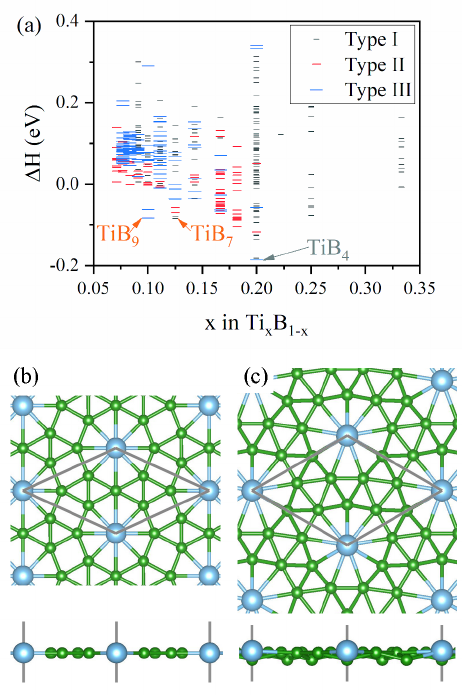}
    \caption{(a) The formation energe of monolayer titanium boride as a function of concentration (x) of titanium. The top and side views of the atomic structure of predicted (b) \ce{TiB7} and (c) \ce{TiB9}.}
    \label{fig:structure}
\end{figure}

As shown in Fig. \ref{fig:how-constructed}, we introduce multi-vacancies (type I, II and III for the \ce{B2}, \ce{B3}and \ce{B4} vacancies) in the boron triangular lattice, where the center of the vacancy is positioned with a single titanium atom. In such cases, titanium atoms are intially embedded in 8, 9, 10-membered rings, respectively. Based on the HNF matrices and structural recognition, we generated the possible candidates of \ce{TiB_x} monolayers and the stable structures were further obtained through the structure relaxation. 

To descirbe the stability of \ce{TiB_x} monolayers, the formation enthalpy($\Delta H$) is calculated as:
\begin{equation}
    \Delta H =(E_{\ce{TiB_x}}-E_{Ti}-xE_{B})/(1+x)
\end{equation}
,  where $E_{TiB_x}$, $E_{Ti}$ and $E_{B}$ are the total energies of one unit cell of the \ce{TiB_x} monolayer, a single titanium atom in hexagonal close-packed (hcp) structure, a single boron atom in $\alpha$-borophene, respectively.

Fig.\ref{fig:structure}(a) shows the $\Delta H$ of \ce{Ti_xB_{1-x}} as a function of titanium concentration. There are two \ce{TiB_4} monolayers with lower  formation energy, which are optimized from type I and III candidates. With titanium concentraiton higher than 0.2, the structures of type I are more stable, while type II and III candidates become more stable when the titanium concentraiton decrease. 

%% describe the structure details
Among the \ce{TiB7} structures, the one with lowest $\Delta H$ (shown in Fig.\ref{fig:structure}(b))
is completely flat. The relaxed lattice constants are a=5.72\AA{} and $\gamma=130.94^{\circ}$. The spacegroup of the structure is Cmmm.
In the structure, each titanium atom is octa-coordinated to boron atoms,
the length of Ti-B bond are 2.148\AA{} , 2.185\AA{}  and 2.373\AA{}  respectively.
The boron atoms forming a network composed of 3,4-membered rings connect to the Ti@\ce{B8} motif.
For the most stable \ce{TiB9} (shown in Fig.\ref{fig:structure}(c)), the relaxed lattice constants of the structure are a=5.83\AA{}, $\gamma=120.0^{\circ}$, with the spacegroup P31m.
Each titanium atom is nona-coordinated to boron atoms, with the Ti-B bond length of 2.216\AA{} , 2.432\AA{}  and 2.430\AA{} respectively.
The boron atoms form a network composed of 3-membered rings connect to the Ti@\ce{B9} motif.

There are 14 different concentrations in the searching sub-space. Besides \ce{TiB7} and \ce{TiB9} as shown in the Fig.\ref{fig:structure}, the most stable configurations of other concentrations are displayed in the ESI. We find that structures with either too high or too low titanium ratio are unstable. In the high titanium ratio conditions, titanium atoms tend to aggregate and the aggregations are then seperated from the boron atoms. In the low titanium ratio configurations, boron atom forms a network structure. And because we didn't consider vacancies in constructing searching candidates, the low titanium structures here are necessarily no more stable than those with vacancies. We can also see from the stable structures of each concentration that \ce{TiB4}, \ce{TiB7}, \ce{Ti2B9} and \ce{TiB5} are totally flat monolayer also known as real two-dimensional structures such as graphene and borophene. In order to forming such kinds of novel structure, the electronic charge need to be avaragely distributed between titanium and borons which require the ratio of titaniums restricted in the suitable range.   

\begin{figure}[htb]
    \centering
    \includegraphics[width=8.3cm]{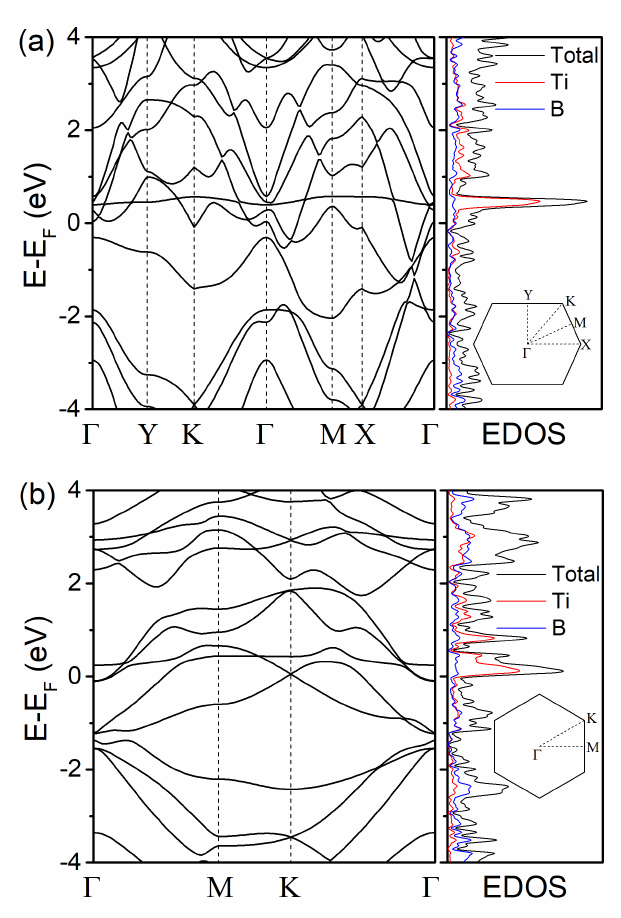}
    \caption{The band structures and electronic density of states of (a) TiB7 and (b) TiB9. The first Brillouin zones are shown in the inset.}
    \label{fig:es}
\end{figure}

Fig.\ref{fig:es} show the electronic properties of the \ce{TiB7} and \ce{TiB9} monolayers, including the projected density of states(PDOS) and band structures. Both the  the \ce{TiB7} and \ce{TiB9} monolayers are metal due to the bands across the Fermi level, which are mainly attributed by the Ti-$d$ states. Comparing to the \ce{TiB4} monolayer\cite{Qu2017}, the $d$ states in these monolayers provide a similar flat band, which increases the density of states near the Fermi level effectively, indicating the possible superconductivity.

\subsection{Superconducticity of  \ce{TiB7} and \ce{TiB9} monolayers} \label{ss:superconducting}

\begin{figure}[h]
    \centering
    \includegraphics[width=8.3cm]{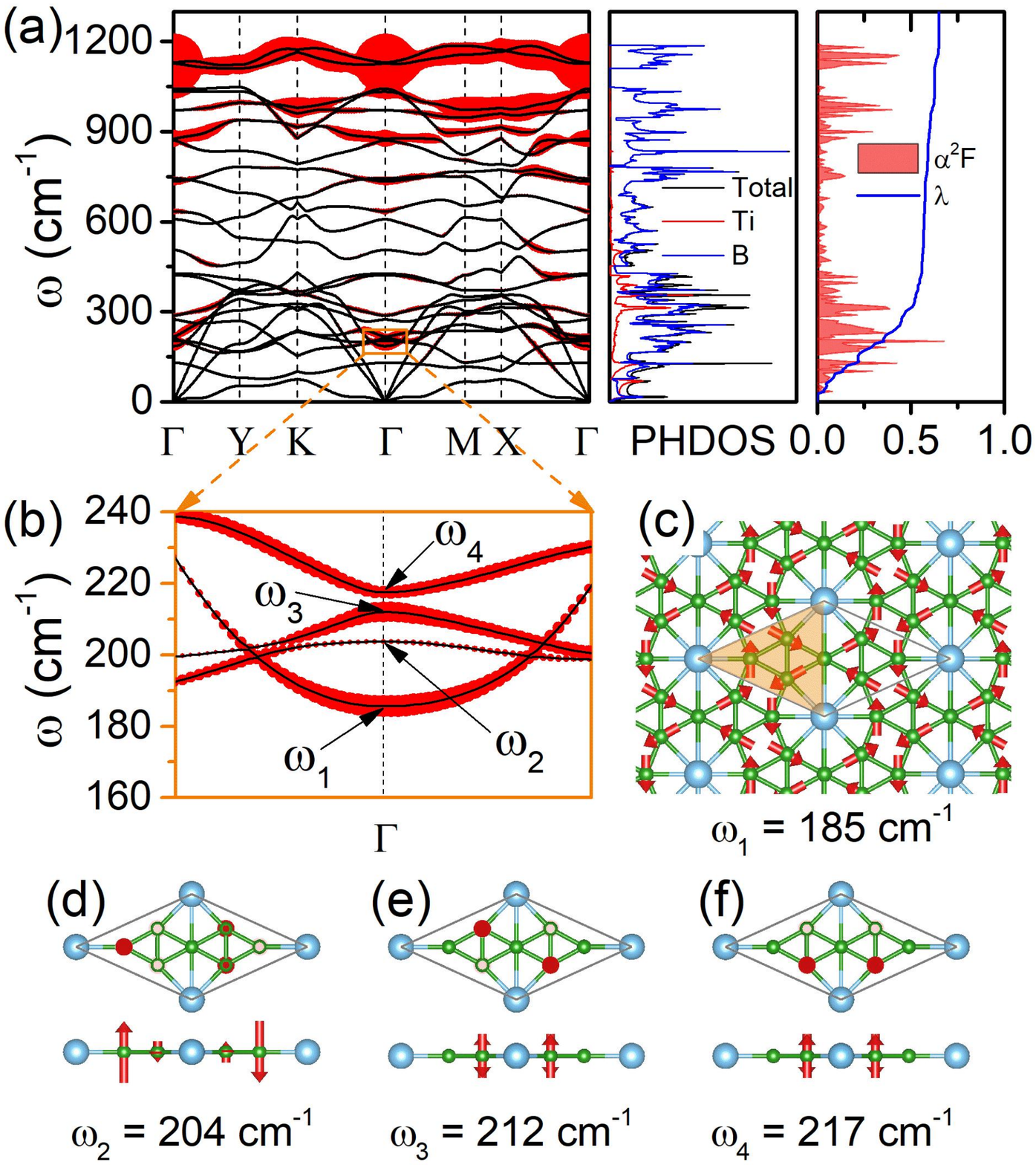}
    \caption{(a) Phonon dispersions with phonon linewidth $\gamma_{qv}$ in red bubble, phonon density of states and Eliashberg function $\alpha^2 F(\omega)$  with $\lambda(\omega)$ of monolayer \ce{TiB7}. (b) A larger view of the rectangular area in (a). (c) A shear mode at $\Gamma$. The filled triangle covers exactly half of a unitcell. (d-f) Some layer breathing modes. The red arrows and their lengths represent the directions and amplitudes of the corresponding vibrational modes.}
    \label{fig:phonon}
\end{figure}

To consider the possible superconductivity, we further studied electron-phonon coupling (EPC) in  the \ce{TiB7} monolayer.
The dynamic stability of \ce{TiB7} was inferred from the phonon spectra without imaginary frequency (see {Fig. \ref{fig:phonon}a}). The EPC constant $\lambda$ was a dimensionless measurement of Eliashberg spectral function $\alpha^2 F(\omega)$,

\begin{equation}
    \lambda = 2 \int \frac{\alpha^2 F(\omega)}{\omega} \mathrm{d}\omega.
\end{equation}

The $\alpha^2 F(\omega)$ was calculated with following equation,

\begin{equation}
    \alpha^2 F(\omega) = 
    \frac{1}{2\pi N (E_F)} \sum_{qv}\delta(\omega-\omega_{qv})\frac{\gamma_{qv}}{\hbar \omega_{qv}}.
\end{equation}

where $N(E_F)$ is the electronic density of states at the Fermi level, and the Dirac $\delta$ function is simulated by a Gaussian function. $\gamma_{qv}$ is the linewidth of phonon mode $v$ at wave vector $q$ .

The phonon dispersions with phonon linewidth $\gamma_{qv}$, phonon density of states (PHDOS), Eliashberg function $\alpha^2 F(\omega)$, and $\lambda(\omega)$ of  monolayer \ce{TiB7} are shown in {Fig. \ref{fig:phonon}a}. The low frequency modes ($<400 cm^{-1}$) induce the main EPC. There is a peak at the frequency of $200 cm^{-1}$ in Eliashberg function $\alpha^2 F(\omega)$. The vibrational modes at $\Gamma$ with the frequencies of around $200 cm^{-1}$ have large phonon linewidth. {Fig. \ref{fig:structure}b} shows a larger view of phonon dispersions with phonon linewidth around $\Gamma$ point with the frequencies ranging from $160$ to $240 cm^{-1}$. The shear mode shown in {Fig. \ref{fig:phonon}c} has a $C_3$ rotation operation. These layer breathing modes shown in {Fig. \ref{fig:phonon}d}, {\ref{fig:phonon}d}, and {\ref{fig:phonon}f} contain rotoreflection, rotational, and mirror symmetry, respectively. The shear and layer breathing modes represent the vibrations perpendicular and parallel to normal\cite{Zhang2013}. The shear mode with a frequency of $185 cm^{-1}$ and the layer breathing modes with frequencies of 212 and $217 cm^{-1}$ have large phonon linewidth. The phonon linewidth of the layer breathing mode at $\Gamma$ point with the frequency of $204 cm^{-1}$ is smaller than those of mentioned three modes, but the phonon branch including this mode is fairly flat around $\Gamma$ point. As a result, a peak appears at the frequency of $200 cm^{-1}$ in Eliashberg function $\alpha^2 F(\omega)$. Meanwhile the mentioned four modes are caused by the vibrations of boron atoms. 

We estimate superconducting transition temperature($T_c$) by using the McMillan-Allen-Dynes parameterized Eliashberg equation \cite{Allen1975},

\begin{equation}
    T_c = \frac{\omega_{\log}}{1.2}
    \exp \left( -\frac{1.04(1+\lambda)}{\lambda-\mu^{\ast}(1+0.62\lambda)} \right),
\end{equation}

where $\omega_{log}$ is the logarithmic averaged phonon frequency

\begin{equation}
    \omega_{\log} = 
    \exp \left( \frac{2}{\lambda} 
    \int \frac{\alpha^2 F(\omega) \log(\omega)}{\omega}\mathrm{d}\omega \right).
\end{equation}

\begin{figure}[h]
    \centering
    \includegraphics[width=8.3cm]{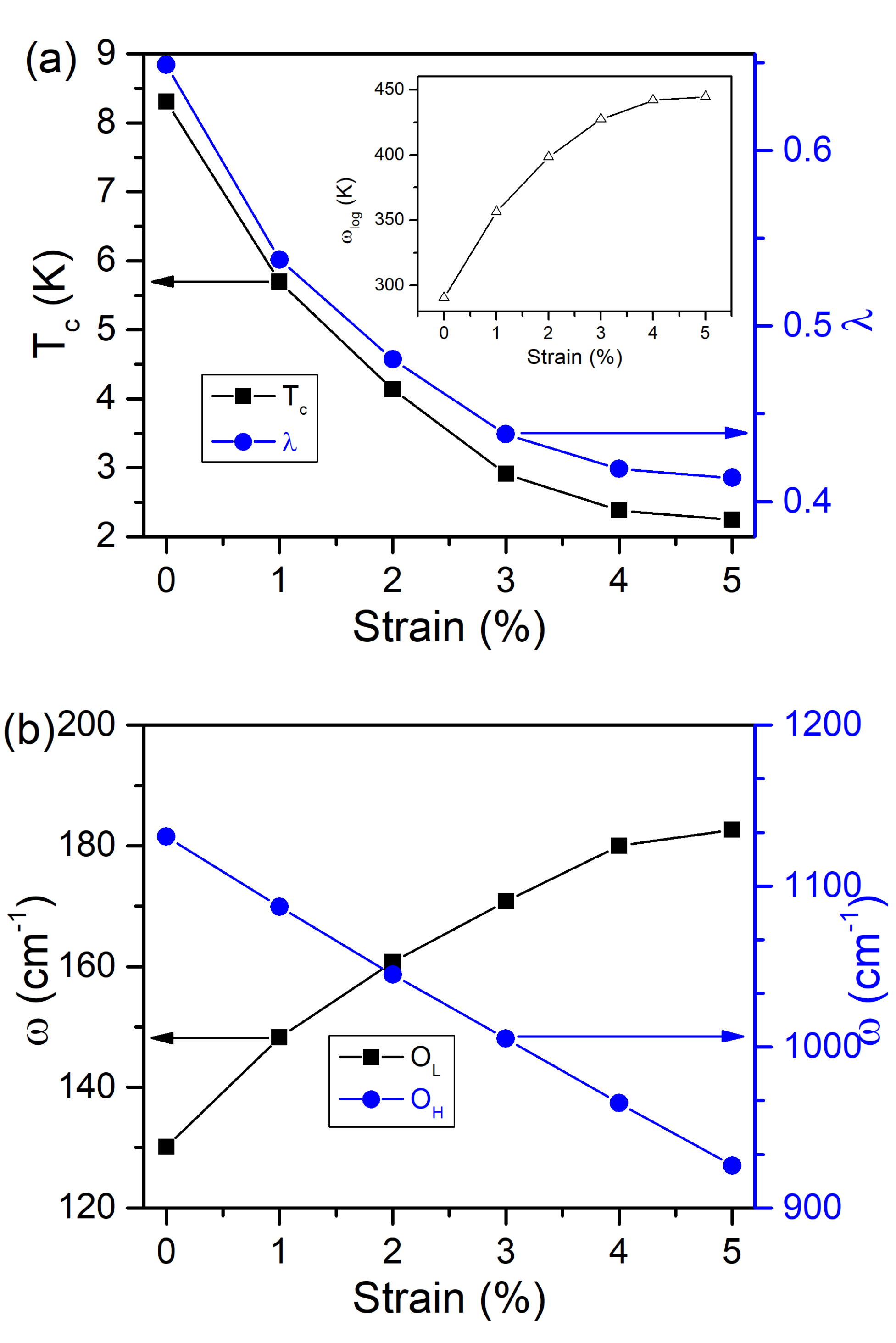}
    \caption{(a) Calculated superconducting transition temperature ($T_c$) and electron-phonon coupling constant ($\lambda$) of monolayer \ce{TiB7} as a function of equibiaxial tensile strain. The evolution of the logarithmic averaged phonon frequency ($\omega_{log}$) is shown in the inset. (b) The lowest ($O_L$) and highest ($O_H$) frequencies of optical phonons at $\Gamma$ as a function of strain.}
    \label{fig:sct}
\end{figure}

Following the previous studies on superconductivity in boron-base materials \cite{Zhao2016,Penev2016,Liao2017,Zhao2018,Yan2020}, we use the Coulomb repulsion pseudopotential $\mu^\ast = 0.1$ to calculate the transition temperature. The estimated $T_c$ of unstrained monolayer \ce{TiB7} reaches 8.3K with $\lambda$ equals to 0.65. In the most stable \ce{B8} sheet ($\alpha$-sheet, $\eta=1/9$ \cite{Tang2007,Yang2008}), the $T_c$ is 3.7K and with $\lambda=0.52$. In the \ce{B7} sheet ($\eta=1/8$), the $T_c$ and $\lambda$ increase to 5.8K and 0.58, respectively \cite{Zhao2016}. So the introduction of titanium atom in \ce{B7} sheet strengthen the electron-phonon coupling and enlarge the transition temperature $T_c$. The two-dimensional \ce{AlB6} is reported to be an intrinsic BCS-type superconductor and the $T_c$ is predicted to be greatly enhanced to 30 K by a equibiaxial tensile strain of 12\% \cite{Song2019}. However, the superconductivity in monolayer \ce{TiB7} is significantly suppressed by tensile strain, as shown in {Fig. \ref{fig:sct}a}. The $T_c$ and $\lambda$ decrease in unison even though the $\omega_{\log}$ increases in sequence. The frequencies of low frequency phonons increase by tensile strain, and the frequencies of high frequency phonons decrease gradually. {Fig. \ref{fig:sct}b} shows the lowest ($O_L$) and highest ($O_H$) frequencies of optical phonons at $\Gamma$ as a function of strain. The details of the evolution of phonon spectra are shown in Fig. S1. Phonon-mediated superconductivity of \ce{B5} sheet ($\beta_{12}$, $\eta=1/6$) is also suppressed by tensile strain \cite{Cheng2017}. In addition, a privious experiment showed that tensile strain suppresses superconductivity in \ce{FeSe} films \cite{Nie2009}. Generally, if the tensile strain suppress the superconductivity of a material, the compressive strain would enhance it, and vice versa, such as the triangular boron monolayer without vacancy ($\delta_6$, $\eta=1/6$) \cite{Xiao2016} and two dimensional \ce{TiS2} \cite{Liao2020}. So we expected the compressive strain could enhance the superconductivity of monolayer \ce{TiB7}. However, imaginary frequencies appear in the phonon spectra when applying the compressive strain of 1\% (see Fig. S2). 

For comparison, we have also studied the \ce{TiB9} monolayer, for which there is no imaginary frequency in the phonon spectra (see Fig. S3). The EPC calculation shows a small $\lambda$ of 0.40. The $T_c$ is predicted to 1.22 K. With  smaller $\lambda$ of 0.10 and 0.27, the recently proposed two dimensional \ce{TiB2} \cite{Zhang2014} and \ce{Ti2B2} \cite{Wang2019} are  not superconductors because of too weak electron-phonon coupling. The most stable monolayer \ce{TiB4}, a completely planar structure \cite{Qu2017}, is also not superconducting due to a small $\lambda$ of 0.18.

\section*{Conclusions}

In summary, we have theoretically investigated a series of planar two dimensional structures of \ce{TiB_x} ($4\leq x \leq 11$)  through high-throughput first-principles calculations. We successfully predicted a superconducting metal-boron monolayer of \ce{TiB7} at the first time, where the $T_c$ is as high as 8K. A new structure of \ce{TiB9} was also found to be a thermodynamic stable monolayer. In our study, the transition metal monolayers are generated with reasonable motifs which can be sumarized as a new approach to quickly search target structures under certain a prior structure constrains. Consequently, the motif based search strategy can be extented and used to explored the new materials with specific structure pattern.
\section*{Conflicts of interest}
There are no conflicts to declare.

\paragraph{Acknowledgements}
This work was supported by Guangdong Natural Science Funds for Distinguished Young Scholars (No. 2014A030306024) and for Doctoral Program (Grant No. 2017A030310086), National Natural Science Foundation of China (No. 11474100, U1601212), the Fundamental Research Funds for the Central Universities(No. 2018ZD46).

%	\newpage
\bibliography{refs}

% \appendix

% \section{Omitted Proof in Section~\ref{sec:examples}}
% \label{app:1}

% \lipsum[7]

\end{document}